\documentclass [prl,twocolumn,longbibliography]{revtex4-2}

\usepackage{graphicx}
\usepackage{subfigure}
\usepackage{amsmath}
\usepackage{amsthm}
\usepackage{amssymb}
\usepackage{hyperref}
\usepackage{xcolor}
\usepackage{textcomp}
\usepackage[normalem]{ulem}
\usepackage{booktabs} 

\hypersetup{colorlinks=true, linkcolor=black, citecolor=black, urlcolor=blue}

\newcommand{\ket}[1]{\left| #1 \right>} % for Dirac bras
\newcommand{\ols}[1]{\mskip.5\thinmuskip\overline{\mskip-.5\thinmuskip {#1} \mskip-.5\thinmuskip}\mskip.5\thinmuskip} % overline short

\begin{document}
\title{Tuning photon-mediated interactions in a multimode cavity: from supersolid to insulating droplets hosting phononic excitations}
\author{Natalia Masalaeva}
\email[Corresponding author: ]{natalia.masalaeva@uibk.ac.at}
\affiliation{Institut f\"ur Theoretische Physik, Universit{\"a}t Innsbruck, A-6020~Innsbruck, Austria}
\author{Helmut Ritsch}
\affiliation{Institut f\"ur Theoretische Physik, Universit{\"a}t Innsbruck, A-6020~Innsbruck, Austria}
\author{Farokh Mivehvar}
\affiliation{Institut f\"ur Theoretische Physik, Universit{\"a}t Innsbruck, A-6020~Innsbruck, Austria}

\begin{abstract}
Ultracold atoms trapped in laser-generated optical lattices serve as a versatile platform for quantum simulations. However, as these lattices are infinitely stiff, they do not allow to emulate phonon degrees of freedom. This restriction can be lifted in emerged optical lattices inside multimode cavities. Motivated by recent experimental progress in multimode cavity QED, we propose a scheme to implement and study supersolid and droplet states with phonon-like lattice excitations by coupling a Bose gas to many longitudinal modes of a ring cavity. The interplay between contact collisional and tunable-range cavity-mediated interactions leads to a rich phase diagram, which includes elastic supersolid as well as insulating droplet phases exhibiting roton-type mode softening for a continuous range of momenta across the superradiant phase transition. The non-trivial dynamic response of the system to a local density perturbation further proves the existence of phonon-like modes.
\end{abstract}

\maketitle

\emph{Introduction.}---A supersolid, one of the most enigmatic states of matter, combines the periodic density modulation of a solid with the dissipationless flow of a superfluid~\cite{Gross1957,Chester1970, Boninsegni2012}. The minimal fundamental requirement for obtaining supersolid is the spontaneous breaking of two continuous symmetries: the internal gauge symmetry which results in superfluidity, and the external spatial translation invariance leading to crystalline order. Despite being predicted to exist in solid ${^{4}\text{He}}$ long ago~\cite{Leggett1970,Thouless1969,Andreev1971}, its experimental observation in helium remains elusive yet~\cite{Kim2012,Chan2013}. 

Ultracold atoms with long-range interactions have proved themselves as an alternative candidate for the realization of the supersolid state. The nature of these interactions can stem from light-mediated interactions~\cite{Munstermann2000,Mottl2012}, or electric or magnetic dipole moments of atoms~\cite{Lu2015,Henkel2010,Cinti2010}. Experimental observations of supersolid properties in Bose-Einstein condensates (BECs) with global light-mediated interactions have been achieved using two crossed linear cavities~\cite{Lonard2017,Leonard2017}, a four-mode ring cavity~\cite{Schuster2020}, bichromatic counterpropagating laser fields~\cite{Dimitrova2017}, and spin-orbit coupling~\cite{Li2017,Bersano2019}. In these experiments, the breaking of two continuous symmetries gives rise to an infinitely degenerate steady-state manifold with a roton-type softened mode at one finite momentum~\cite{Leonard2017}. Nevertheless, due to the global nature of the interactions, the resultant supersolid turns out to be infinitely stiff and impervious to phononic excitations that determine thermodynamic properties of real materials. On the other hand, supersolid states realized in dipolar quantum gases overcome this shortcoming~\cite{Tanzi2019,Bttcher2019,Chomaz2019,Guo2019Goldstone,Tanzi2019super,Natale2019}: The  finite-range, momentum-dependent dipole-dipole interactions between atoms result in an excitation spectrum exhibiting energy minima at continuum of finite momenta~\cite{Santos2003,Chomaz2018,Petter2019}---roton mode~\cite{Landau1941}---similar to the one observed in superfluid helium~\cite{Yarnell1958}. The simultaneous observation of periodic density modulation and global phase coherence~\cite{Tanzi2019,Bttcher2019,Chomaz2019}, together with sound modes~\cite{Guo2019Goldstone,Tanzi2019super,Natale2019}, confirmed the genuine supersolid nature of these states.

\begin{figure}[b!]
\centering
\includegraphics [width=0.98\linewidth]{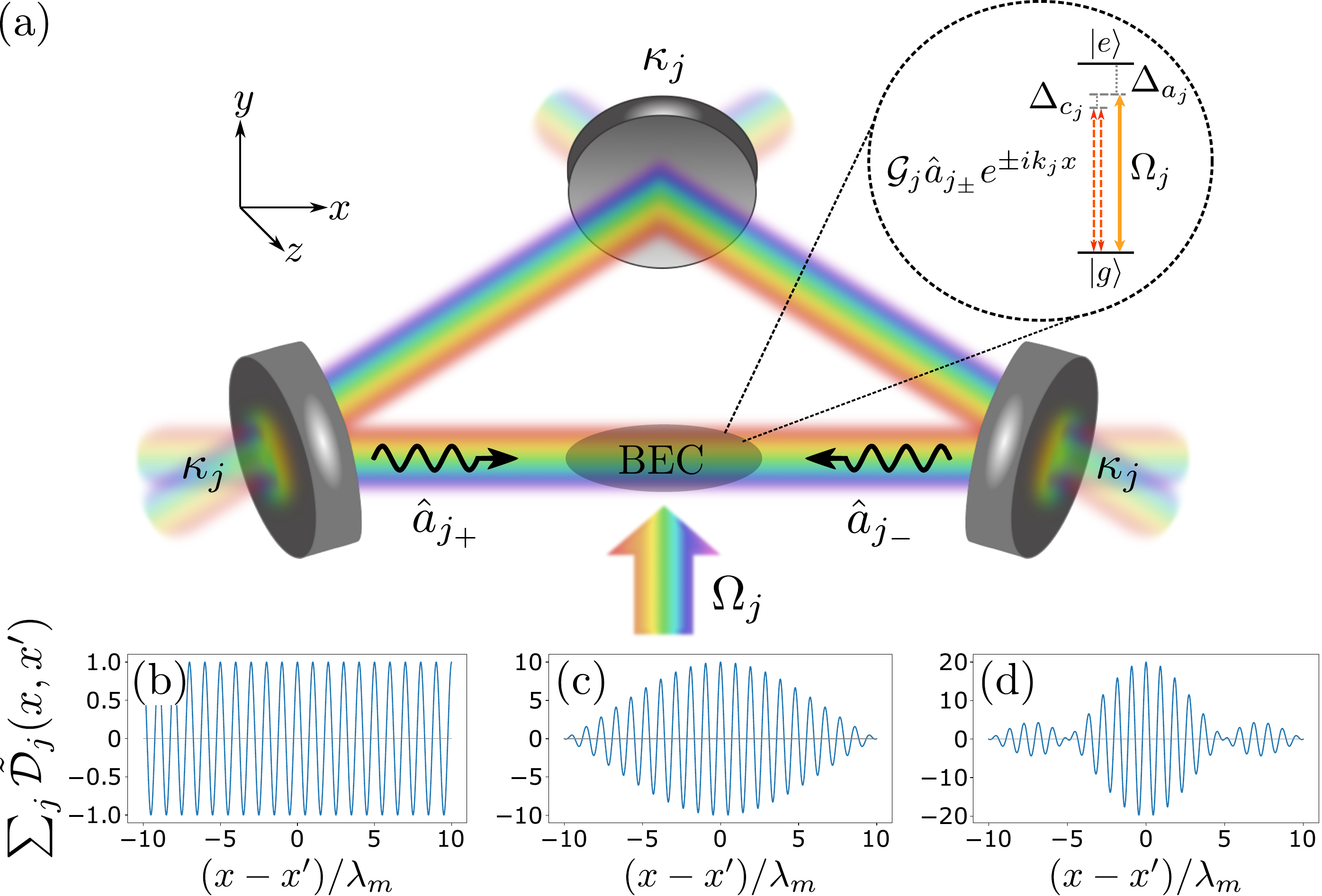}
\caption{(a) Sketch of the system: A BEC is coupled to {$N_c$} longitudinal running-wave modes ${\hat a}_{j_{\pm}}$ of a ring cavity and transversely illuminated by $N_c$ lasers with Rabi frequencies $\Omega_j$. 
%The inset shows the atom-photon coupling scheme. 
(b)-(d) The dimensionless cavity-mediated interaction potential $\sum_j\mathcal{\tilde{D}}_j(x,x')$ as a function of the relative interatomic distance for different number of involved modes: (b) $N_c = 1$, (c) $N_c = 10$, and (d) $N_c = 20$.}
\label{fig:Scheme}
\end{figure}

\begin{figure*}[t!]
\centering
\includegraphics [width=0.98\textwidth]{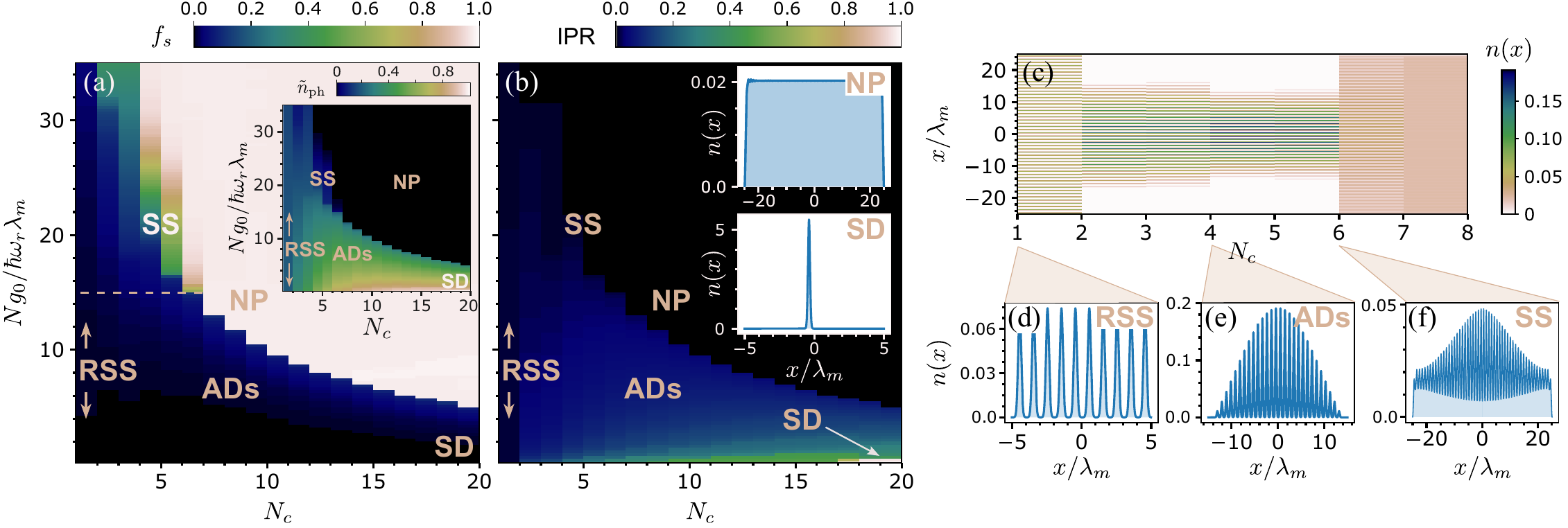}
\caption{Phase diagram of the system. The superfluid fraction $f_s$ (a) and the inverse participation ratio $\text{IPR}$ (b) as functions of the contact-interaction strength $N g_0/\hbar\omega_r\lambda_m$ and the number of involved modes $N_c$ related to the effective range of the cavity-mediated interactions. Five regimes are identified in the phase diagrams: rigid supersolid (RSS), elastic supersolid (SS), array of droplets (ADs), single droplet (SD), and a normal phase (NP). The inset in (a) shows the rescaled total photon number $\tilde{n}_{\rm ph}$ in all modes in the same parameter plane. (c) A horizontal cut through the phase diagram at $N g_0 = 15\,\hbar\omega_r\lambda_m$ [the brown dashed line in panel (a)] shows the corresponding change in the atomic density $n(x)$ for a varying number of involved modes. (d)--(f) Exemplary density profiles for RSS at $N_c = 1$ (d), for ADs at $N_c = 4$ (e), and for SS at $N_c = 6$ (f). The insets in (b) display typical atomic density profiles for NP ($N g_0 = 15\,\hbar\omega_r\lambda_m$, $N_c = 13$) and for SD ($N g_0 = 0.5\,\hbar\omega_r\lambda_m$, $N_c = 20$). The parameters are set to $(\sqrt{N}\eta,N U,\kappa) = (3.2/\sqrt{N_c},-1,1)\,\omega_r$ and $\lambda_m = L_{\text{cav}}/m$ with $m = 50$ being the lowest mode in the series. Here, $\Delta_{c_j}$ vary from $-10\,\omega_r$ to $-4\,\omega_r$, as shown in the inset of Fig.~\ref{fig:Dispersion}(a) for $N_c = 20$.}
\label{fig:Ph_diagram}
\end{figure*}

\begin{figure}[t!]
\centering
\includegraphics [width=0.9\linewidth]{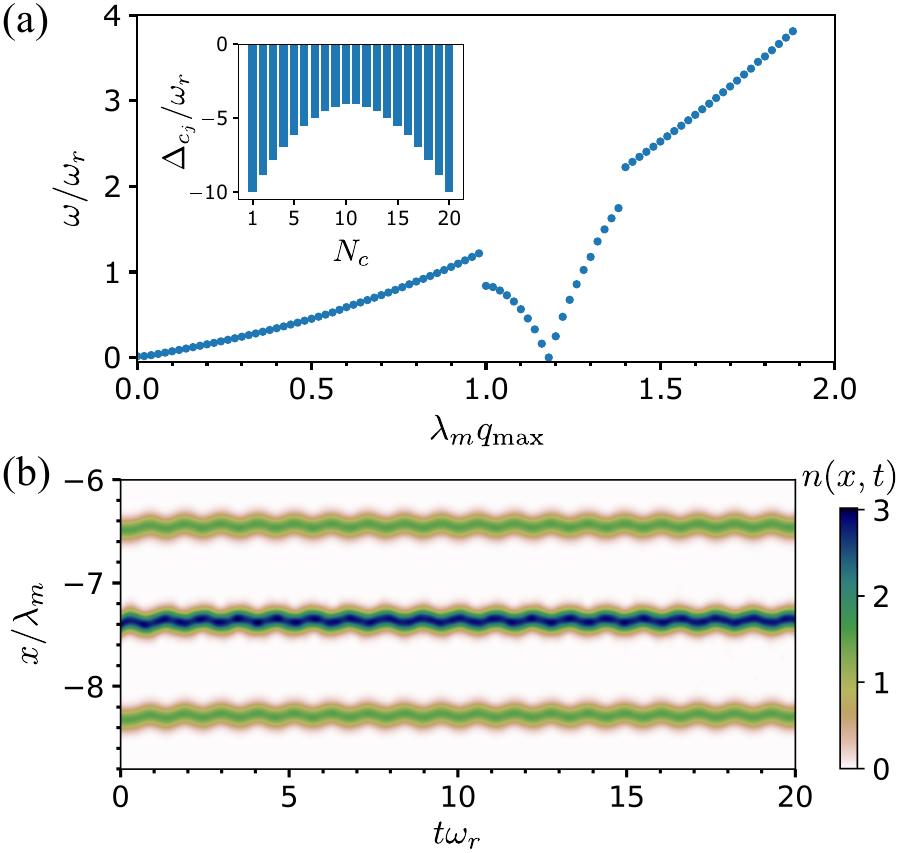}
\caption{Excitation spectrum and real-time dynamics. (a) Excitation spectrum in NP slightly below the threshold with $\sqrt{N}\eta = 6.4/\sqrt{N_c}\,\omega_r$ where $N_c = 20$ and $N g_0 = 15\,\hbar\omega_r\lambda_m$. Here, $q_{\text{max}}$ is the dominant momentum component in each Bogoliubov eigenvector; see the main text. The momentum discretization $\Delta q$ is related to the system size via $\Delta q = 1/50\lambda_m$. The inset shows the distribution of the cavity detunings $\{\Delta_{c_j}\}$. (b) Time evolution of the atomic density $n_{\rm pert}(x,t)$ after the initial kick of the central droplet with velocity $v = 5/\lambda_m$. The initial steady state is for the parameters $\sqrt{N}\eta = 3.2/\sqrt{N_c}$, $N g_0 = 0.5\,\hbar\omega_r\lambda_m$, and $N_c = 10$. The other parameters are the same as in Fig.~\ref{fig:Ph_diagram}.}
\label{fig:Dispersion}
\end{figure}

An alternative route for realizing non-stiff supersolid is based on multimode cavity QED~\cite{Gopalakrishnan2009,Gopalakrishnan2010}, where most notably a lattice phononic mode was recently observed in a Bose gas inside a confocal multimode cavity as a major step towards non-stiff cavity-QED supersolid~\cite{Guo2021}. Motivated by the recent progress in this direction~\cite{Mivehvar2021,Wickenbrock2013,Kramer2014,Vaidya2018,Keller2018,Guo2019Sign,Guo2019Emergent,Colella2022,Karpov2023,Uhrich2023}, in this Letter we demonstrate a concrete scheme to implement and study nonequilibrium supersolid states hosting phononic lattice excitations via coupling a BEC into many \emph{longitudinal} modes of a ring cavity; see Fig.~\ref{fig:Scheme}(a). Such a system with a very few counterpropagating running modes forms a stiff supersolid~\cite{Gopalakrishnan2010,Mivehvar2018,Schuster2020}. However, we show that by populating many longitudinal modes, localized photon wavepackets---supermodes as introduced in the context of confocal multimode cavities~\cite{Kollr2017}---are formed leading to finite-range interatomic interactions [Fig.~\ref{fig:Scheme}(c) and (d)] and roton mode softening for a continuum of finite momenta [Fig.~\ref{fig:Dispersion}(a)]. Depending on the interplay between these tunable-range cavity-mediated and contact collisional interactions, the system is in either a supersolid or an insulating droplet state (Fig.~\ref{fig:Ph_diagram}), both capable of supporting lattice phonons. This is corroborated by the dynamic response of the system to a local density perturbation [Fig.~\ref{fig:Dispersion}(b)]. 

\emph{Model.}---Consider two-level bosonic atoms inside a ring cavity strongly confined along the cavity axis by a box potential $V_{\rm box}(x)$ of length $L$. The atomic transition $\ket g\leftrightarrow\ket e$ is off-resonantly coupled to $N_c$ pairs of longitudinal, counterpropagating running-wave modes of the cavity, as depicted in Fig.~\ref{fig:Scheme}(a), with coupling strengths ${\hat a}_{j_{\pm}}\mathcal{G}_{j\pm}(x) = {\hat a}_{j_{\pm}} \mathcal{G}_j e^{\pm i k_j x}$, where ${\hat a}_{j_{\pm}}$ are the photonic annihilation operators and $\pm k_j = \pm \omega_j/c$ are the wave numbers of the $j$th pair of  degenerate cavity modes traversing to right/left, respectively. Here, $j = m,\ldots,m + N_c - 1$ where $m$ determines the lowest involved mode with wavelength $\lambda_m = L_{\text{cav}}/m$ and $L_{\text{cav}}$ being the cavity length. 
%The cavity modes are initially in the vacuum state. 
The atoms are further illuminated from the side by $N_c$ off-resonant external pump lasers inducing the same atomic transition with corresponding Rabi frequencies $\Omega_j$. The pump and cavity frequencies, $\omega_{p_{j}}$ and $\omega_j$, are assumed to be far red-detuned from the atomic transition frequency $\omega_a$, but are close to the resonance with each other. In particular, the $j$th laser is in near-resonance with only the $j$th pair of the cavity modes.

In the dispersive regime, i.e., for large average relative atomic detuning $\ols\Delta_a\equiv\sum_j\omega_{p_j}/N_c - \omega_a$, the excited state $\ket e$ has a negligible population and can be adiabatically eliminated~\cite{Ritsch2013}. This yields an effective many-body Hamiltonian for the ground state $\ket{g}$ as detailed in supplemental material (SM)~\cite{SM2023},
\begin{align} \label{eq:H_mb}
{\hat H}_\text{eff} &= \int\hat\psi^\dagger(x)\hat{\mathcal{H}}^{(1)}_{\text{eff}}\hat\psi(x)dx -\hbar\sum_{j = m}^{m + N_c-1}\Delta_{c_{j}}\big({\hat a}^\dagger_{j_{+}}{\hat a}_{j_{+}} \nonumber\\
&+ {\hat a}^\dagger_{j_{-}}{\hat a}_{j_{-}}\big) +g_0\int\hat\psi^\dagger(x)\hat\psi^\dagger(x)\hat\psi(x)\hat\psi(x)dx,   
\end{align}
where $\hat{\psi}(x)$ is the atomic bosonic field operator and $\Delta_{c_{j}}\equiv\omega_{p_{j}} - \omega_j$ denotes the relative frequency between a pump laser and the corresponding cavity modes. The effective single-particle Hamiltonian density reads
\begin{align} \label{eq:H_sp}
\hat{\mathcal{H}}^{(1)}_{\text{eff}} &= -\frac{\hbar^2}{2M}\frac{\partial ^2}{\partial x^2} 
+V_{\rm box}(x)
+ \hbar\sum_j \Big[U_j\big({\hat a}^\dagger_{j_{+}}{\hat a}_{j_{+}} + {\hat a}^\dagger_{j_{-}}{\hat a}_{j_{-}} \nonumber\\
& + {\hat a}^\dagger_{j_{+}}{\hat a}_{j_{-}}e^{-2i k_j x} + {\hat a}^\dagger_{j_{-}}{\hat a}_{j_{+}}e^{2i k_j x}\big) \nonumber\\
& + \eta_j\big({\hat a}_{j_{+}}e^{i k_j x}  + {\hat a}_{j_{-}}e^{-i k_j x} + \text{H.c.}\big)\Big]. 
\end{align}
Here, we have introduced $U_j\equiv \mathcal{G}_j^2/\ols\Delta_a$ and $\eta_j\equiv\mathcal{G}_j\Omega_j/\ols\Delta_a$. In order to keep the number of variables tractable, in the following we set $U=U_j$ and $\eta=\eta_j$ for all $j$. Note that we have neglected cross-scattering terms between different pairs of cavity modes and pumps since such scatterings are energetically not favored (i.e., they oscillate very fast~\cite{SM2023}). The two-body interactions between the atoms are represented by the local contact interaction with strength $g_0$.

The system possesses two continuous $U(1)$ symmetries corresponding to the superfluid gauge freedom $\hat{\psi}\to\hat{\psi}e^{i\theta}$, and the invariance under the `composite' transformation $x\to x+\Delta x$ and $\hat{a}_{j\pm}\to\hat{a}_{j\pm}e^{\mp ik_j\Delta x}$. The former $U(1)$ symmetry is broken by the condensate wave function at the BEC transition, while the latter is broken at the onset of the superradiant phase transition, signaling the formation of an (at least, minimal) supersolid state in this regime.

\emph{Cavity-mediated tunable-range interactions.}---The exchange of cavity photons by the atoms results in an effective interatomic interaction~\cite{Maschler2008}. In the fast cavity-dynamics regime the cavity fields can be adiabatically eliminated to obtain a compact form for the resultant interatomic potential~\cite{SM2023},
\begin{equation} \label{cmp}
\hat{\mathcal D}(x,x') = \sum_j\frac{4\eta^2 \text{Re}(\hat{\delta}_{c_j})}{|\hat{\delta}_{c_j}|^2}\mathcal{\tilde{D}}_j(x,x'),
\end{equation}
where $\hat{\delta}_{c_j} \equiv \Delta_{c_j} - U\hat{N} + i\kappa$ with $\hat{N} = \int \hat{n}(x) dx $ being the particle-number operator and $\kappa=\kappa_j$ the cavity-field decay rate for all $j$. The spatial dependence of the density-density interaction~\eqref{cmp} is determined by the dimensionless functions $\mathcal{\tilde{D}}_j(x,x') = \cos{[k_j(x-x')]}$, shown in Figs.~\ref{fig:Scheme}(b)-\ref{fig:Scheme}(d) for the case when all the coefficients in Eq.~\eqref{cmp} are equal. As expected for one pair of counterpropagating running modes cavity-mediated interactions are global and periodic in space; see Fig.~\ref{fig:Scheme}(b). However, by increasing the number of involved modes the interaction potential acquires a global minimum and a decaying envelope as shown in Fig.~\ref{fig:Scheme}(d), similar to a multi-mode confocal cavity~\cite{Vaidya2018}, indicating the \emph{finite range} of the interactions. This can be understood in terms of supermodes: cavity modes with different wave numbers can be in phase and hence interfere constructively with each other only in a confined region, thus forming localized photon wavepackets. As will be shown in the following, the interplay between these cavity-mediated tunable-range interactions and the repulsive collisional interactions results in rich physics. 

\emph{Mean-field phase diagram.}---In the thermodynamic limit, where the mean-field description becomes exact~\cite{Piazza2013}, the atomic and cavity-field operators are replaced by their corresponding quantum averages, $\alpha_{j_\pm} (t) = \langle {\hat a}_{j_\pm}(t)\rangle$ and $\psi(x,t) = \langle\hat\psi(x,t)\rangle$. The system is then described by a set of coupled nonlinear equations
\begin{align} \label{eq:mf_eq}
i\frac{\partial}{\partial t} \alpha_{j_\pm} & = -\delta_{c_j}\alpha_{j_\pm} + U\mathcal{N}_{\pm 2}^{(j)}\alpha_{j_\mp} + \eta\mathcal{N}_{\pm 1}^{(j)},\nonumber\\
i\hbar\frac{\partial}{\partial t} \psi(x) & = \left[\mathcal{H}^{(1)}_{\text{eff}}+g_0n(x)\right]\psi(x),
\end{align}
where $\mathcal{H}^{(1)}_{\text{eff}}(\{\alpha_{j\pm}\})$ is the mean-field Hamiltonian corresponding to Eq.~\eqref{eq:H_sp}, and $\mathcal{N}_{\pm 1}^{(j)} = \int n(x)e^{\mp ik_j x}dx$ and $\mathcal{N}_{\pm 2}^{(j)} = \int n(x)e^{\mp 2ik_j x}dx$ for the shorthand.

We find steady states of the cavity-field amplitudes $\partial_t\alpha_{j\pm} = 0$ and the condensate wave function $i\hbar \partial_t\psi = \mu\psi$, with $\mu$ being the chemical potential, by self-consistently solving Eq.~\eqref{eq:mf_eq} on a box of length $L=50\lambda_m$, where $\lambda_m$ is the largest wavelength corresponding to the lowest mode. The superfluidity can roughly be characterized by the superfluid fraction~\cite{Leggett1970,Sepulveda2008}, $f_s = [l^2/\int_{\rm uc}|\psi(x)|^2 dx][\int_{\rm uc}{|\psi(x)|^{-2}}dx]^{-1},$ where $l$ is the length of a unit cell in the bulk. For a uniform state $f_s = 1$, while the density modulation reduces the value of the superfluid fraction, with $f_s = 0$ corresponding to states with no superfluid properties. To characterize the degree of localization of a state, we calculate the inverse participation ratio, $\text{IPR} = \int n^2(x) dx/ [\int n(x) dx]^2$. It takes a maximum value for a completely localized state and smaller values for extended states. The crystalline order can be quantified using the density contrast $\mathcal{C}=
({n_{\text{max}} - n_{\text{min}}})/({n_{\text{max}} + n_{\text{min}}}),$
where $n_{\text{max}}$ ($n_{\text{min}}$) is the maximum (minimum) of the atomic density in the bulk.

Figures~\ref{fig:Ph_diagram}(a)--(b) show the mean-field phase diagram of the system in the parameter plane of the strength $g_0$ of the collisional contact interactions and the number $N_c$ of involved cavity modes, where the latter determines the effective range of the cavity-mediated interactions [see Figs.~\ref{fig:Scheme}(b)-\ref{fig:Scheme}(d)]. In order to \emph{approximately} keep the total energy flux into the system constant, we scale the pump strength with the square root of the number of involved cavity modes, that is, $\eta/\sqrt{N_c}$. This is to say that, by increasing $N_c$ and using the scaling $\eta/\sqrt{N_c}$ we approximately keep the strength of the cavity-mediated interaction potential~\eqref{cmp} constant and only tune its effective range. The superfluid fraction $f_s$ [Fig.~\ref{fig:Ph_diagram}(a)], the inverse participation ratio $\text{IPR}$ [Fig.~\ref{fig:Ph_diagram}(b)], the density contrast $\mathcal{C}$ (see SM~\cite{SM2023}), and the total rescaled photon number $\tilde{n}_{\rm ph}=\sum_j(|\alpha_{j+}|^2+|\alpha_{j-}|^2)/N$ with $N$ being the total atom number [inset of Fig.~\ref{fig:Ph_diagram}(a)] reveal five phases: A normal phase (NP), rigid supersolid (RSS), elastic supersolid (SS), an array of droplets (ADs), and a single droplet (SD) state. 

We note that the transitions from NP into RSS and SS are second order, while from NP into ADs and SD are first order. That is, even though $N_c$ changes discretely, the system variables such as $f_s$, IPR, $\mathcal{C}$, and $\tilde{n}_{\text{ph}}$ change smoothly~\cite{SM2023}. In Figs.~\ref{fig:Ph_diagram}(c)-(f) we illustrate the change of the atomic density $n(x)$ for a fixed contact-interaction strength $N g_0 = 15\,\hbar\omega_r\lambda_m$, where $\omega_r = \hbar k_m^2/2M$ is the recoil frequency, and a varying number of involved cavity modes $N_c$, i.e., along the dashed line shown in Figs.~\ref{fig:Ph_diagram}(a); see also SM for other cuts and additional phase diagrams in the $\{\eta,g_0\}$--parameter plane for fixed $N_c$~\cite{SM2023}. For $N_c = 1$, that is, for only two counterpropagating running modes, the emerging optical potential inside the cavity is $\lambda_m$-periodic [see Fig.~\ref{fig:Scheme}(b)] and the BEC forms a RSS~\cite{Mivehvar2018} as depicted in Fig.~\ref{fig:Ph_diagram}(d) in $10\lambda_m$ for the clarity of the presentation. By including more modes, $N_c > 1$, the emergent optical potential acquires a global minimum; see Figs.~\ref{fig:Scheme}(c) and (d). The position of this global minimum is fixed spontaneously via breaking the `composite' continuous $U(1)$ symmetry of the system. The atomic distribution is ADs in small $N_c$ and crosses over to SS at larger $N_c$; see Figs.~\ref{fig:Ph_diagram}(e) and (f). However, at even larger $N_c$ the system transits into NP and the atomic distribution becomes uniform [except at the edges; see the inset of Fig.~\ref{fig:Ph_diagram}(b)]. This is due to the fact that the scaling $\eta/\sqrt{N_c}$ keeps the effective pumping strength only approximately constant. At large $N_c$, the rescaled pumping strength falls bellow the superradiant threshold.

Let us note that for weak contact interactions and a large number of populated modes, the condensate collapses into the global minimum of the cavity potential and becomes a SD characterized by $\text{IPR} \sim 1$ as shown in the inset of Fig.~\ref{fig:Ph_diagram}(b). This is accompanied by the intense, coherent scattering of pump photons into the cavity and an increase of photon number in all modes; see the inset of Fig.~\ref{fig:Ph_diagram}(a).

\emph{Excitation spectrum and real-time dynamics.}---It is worth noting that the presence of both density modulation and global phase coherence does not serve as conclusive evidence for the supersolid nature of a state~\cite{Pitaevskii2016}. This can be verified unambiguously by dynamical properties of the state. Therefore, here we first calculate the collective excitation spectrum and then study real-time vibrational dynamic after a kick, in order to shed light on the elasticity and phononic modes of the system. 

The tendency of the system to crystallize can be foretold by the existence of the roton mode in the excitation spectrum~\cite{Nozires2004,Kirzhnits1971,Pomeau1994}. To compute the excitation spectrum, we include quantum fluctuations above the mean-field solutions for the condensate wave function, $\delta\psi(x)$, and the cavity-filed amplitudes, $\delta\alpha_{j\pm}$. We linearize Eq.~\eqref{eq:mf_eq} by keeping only linear terms in the fluctuations and obtain Bogoliubov-type equations which can be recast in a compact matrix form~\cite{Horak2001,Nagy2008,Mivehvar2018,SM2023},
\begin{equation} \label{Bogmat}
 \omega \mathbf{f} = \mathbf{M}_B \mathbf{f}, 
\end{equation}
where $\mathbf{f}$ is a vector of the fluctuations and $\mathbf{M}_B$ is a non-Hermitian matrix given explicitly in SM~\cite{SM2023}; see SM for the whole details. The collective excitation spectrum $\omega$ is the solution of this eigenvalue problem. 

To obtain the dispersion relation $\omega(q)$, we numerically diagonalize the Bogoliubov matrix $\mathbf{M}_B$ in position space and then Fourier transform the eigenvectors. Subsequently, we arrange eigenvalues $\omega$ according to the dominant momentum component $q_{\text{max}}$ of their corresponding eigenvectors~\cite{Ostermann2016}. The resultant excitation spectrum is shown in Fig.~\ref{fig:Dispersion}(a) and possesses a roton-type softened mode for a continuum of finite momenta, similar to the one obtained in dipolar quantum gases~\cite{Chomaz2018,Petter2019} and multimode confocal cavities~\cite{Guo2021}. The energy gap is closed across the superradiant phase transition for the momentum corresponding to the cavity wave number $k_j$ with the smallest absolute-value detuning $\Delta_{c_j}$; see inset of Fig.~\ref{fig:Dispersion}(a). In contrast to one-mode cavities, where the roton softening occurs strictly only for one momentum due to the infinite range of cavity-mediated interactions~\cite{Leonard2017}, our system has a broad roton softening caused by finite-range photon-mediated interactions. The softened roton is the Goldstone mode of the broken `composite' $U(1)$ symmetry. This roton branch defines the elastic properties of the crystalline states~\cite{Guo2021}, as we show in the following. Before continuing, it is, however, worth noting that the range of the cavity-mediated interactions~\eqref{cmp} and hence the form of the roton minimum can also be tuned by properly choosing the cavity detunings $\{\Delta_{c_j}\}$, in addition to the number $N_c$ of cavity modes.

To probe the response of the system to a local perturbation and reveal phononic excitations, we kick the system in the ADs phase, consisting of three droplets, via the protocol $\psi_{\rm pert}(x)|_{t = 0} = \psi(x) e^{iv\mathcal{F}(x)x}$. Here, $\psi(x)$ is the steady state of the system, $v$ is the velocity of the kick, and $\mathcal{F}(x)$ is a step-like function, which allows applying the kick only to a specific (i.e., central) droplet. The subsequent nonequilibrium dynamics of the atomic density $n(x)$ obtained from Eq.~\eqref{eq:mf_eq} is depicted in Fig.~\ref{fig:Dispersion}(b). The outer droplets demonstrate in-phase oscillations, while the central droplet vibrates out-of-phase with respect to the outer ones, similar to the prediction for a dipolar gas~\cite{Mukherjee2023}. This is in sharp contrast to the uniform center-of-mass motion of a whole RSS inside a ring cavity under a gravitational perturbation~\cite{Gietka2019Supersolid}. The nontrivial elastic response of the superradiant lattice to the local density perturbation in our system heralds the presence of phononic modes~\cite{Mishra2023}, and it is consistent with the excitation spectrum in Fig.~\ref{fig:Dispersion}(a). 

\emph{Conclusions}---We have shown how a variety of different exotic quantum phases, including supersolid and droplet arrays, can be implemented in a driven BEC inside a multimode cavity by tuning the range of the photon-mediated interactions. The resulting phase diagram resembles ones obtained with dipolar quantum gases~\cite{Tanzi2019,Bttcher2019,Chomaz2019,Blakie2020,Smith2023}, offering an alternative route to exploring many-body physics with tunable-range photon-mediated interactions. In addition, the monitoring of leaked cavity fields provides build-in non-destructive access to resultant phases in real-time. Our model can be implemented in state-of-the-art experiments~\cite{Johnson2019}; see SM for a concrete discussion~\cite{SM2023}.

We further note that in similar setups with significantly closer mode spacing, cross-scattering terms between different cavity modes and pumps start to play a role and mediate interactions between different softened roton modes. In this case, only one pump laser tuned to the central pair of modes would probably be sufficient, which simplifies the experimental realization significantly. Compared to a rigid supersolid~\cite{Gietka2019Supersolid}, our setup should provide improved possibilities for quantum sensing applications thanks to its larger multimodal deformability and frequency range.    

\begin{acknowledgments}
F.\,M.~acknowledges financial supports from the Stand-alone project P~35891-N of the Austrian Science Fund (FWF),
the Tyrolean Science Promotion Fund (TWF) of the State of Tyrol, and the ESQ Discovery Grant of the Austrian Academy of Sciences (\"OAW). F.\,M. and N.\,M.~ are grateful to the Theodor K\"{o}rner Fund for the Promotion of Science and Art for their support.
\end{acknowledgments}

\end{document}

\title{Supplemental material: Tuning photon-mediated interactions in a multimode cavity: from supersolid to insulating droplets hosting phononic excitations}
\author{Natalia Masalaeva}
\email[Corresponding author: ]{natalia.masalaeva@uibk.ac.at}
\affiliation{Institut f\"ur Theoretische Physik, Universit{\"a}t Innsbruck, A-6020~Innsbruck, Austria}
\author{Helmut Ritsch}
\affiliation{Institut f\"ur Theoretische Physik, Universit{\"a}t Innsbruck, A-6020~Innsbruck, Austria}
\author{Farokh Mivehvar}
\affiliation{Institut f\"ur Theoretische Physik, Universit{\"a}t Innsbruck, A-6020~Innsbruck, Austria}

\maketitle

\setcounter{equation}{0}
\setcounter{figure}{0}
\renewcommand{\theequation}{S\arabic{equation}}
\renewcommand{\thefigure}{S\arabic{figure}}
\renewcommand{\bibnumfmt}[1]{[S#1]}
\renewcommand{\citenumfont}[1]{S#1}

\section{SUPPLEMENTAL MATERIAL}
Here we present the details of the adiabatic elimination of the atomic excited state [Eq.~(2) in the main text] and the derivation of the effective cavity-mediated potential [Eq.~(3) in the main text]. Then we provide an additional phase diagram for the density contrast, cuts through the phase diagrams in the main text to illustrate the character of the phase transitions, and additional phase diagrams for a fixed number of involved modes. Moreover, we derive the linearized equations and the Bogoliubov matrix, and, finally, we discuss the possible experimental realization of the suggested setup.

\subsection{Adiabatic elimination of the atomic excited state}

Within the dipole and rotating-wave approximations, the single-particle Hamiltonian for the system presented in the main text reads
\begin{equation} \label{app_h_td} 
    \hat{\mathcal{H}}^{(1)} = -\frac{\hbar^2}{2M}\frac{\partial^2}{\partial x^2} I_{2\times 2} + \hbar\omega_e \hat\sigma_{ee} + \hbar\sum_j\Big\{\big[e^{-i\omega_{p_j}t}\Omega_j + \mathcal{G}_j( \hat{a}_{j_+}e^{i k_j x} + \hat{a}_{j_-}e^{-i k_j x})\big]\hat{\sigma}_{eg} + \text{H.c.}\Big\} + \hbar\sum_j\omega_{c_j}\big({\hat a}^\dagger_{j_+}{\hat a}_{j_+} + {\hat a}^\dagger_{j_-}{\hat a}_{j_-}\big),
\end{equation}
where $m$ is the atomic mass, $\hat\sigma_{\tau\tau'} = |\tau\rangle\langle\tau'|$ is the atomic transition operator with $\tau = \{g,e\}$. The single-particle Hamiltonian density~\eqref{app_h_td} can be brought into the time-independent form 
\begin{align} \label{app_h} 
\hat{\tilde{\mathcal{H}}}^{(1)} & = \hat U\hat H{\hat U}^\dagger + i\hbar(\partial_t\hat U){\hat U}^\dagger \nonumber\\
    & = -\frac{\hbar^2}{2M}\frac{\partial^2}{\partial x^2} -\hbar\ols\Delta_a\hat\sigma_{ee} + \hbar\sum_j\Big\{\big[\Omega_j + \mathcal{G}_j(\hat{a}_{j_+}e^{i k_j x} + \hat{a}_{j_-}e^{-i k_j x})\big]\hat{\sigma}_{eg}e^{i\delta_{p_j} t} + \text{H.c.}\Big\} - \hbar\sum_j\Delta_{c_j}\big({\hat a}^\dagger_{j_+}{\hat a}_{j_+} + {\hat a}^\dagger_{j_-}{\hat a}_{j_-}\big),
\end{align}
using the unitary operator
\begin{equation} 
    \hat U = \text{exp}\Big(i\ols\omega_p\hat\sigma_{ee}t + it\sum_j\omega_{p_j}\big[{\hat a}^\dagger_{j_+}{\hat a}_{j_+} + {\hat a}^\dagger_{j_-}{\hat a}_{j_-}\big]\Big).
\end{equation}
Here we have defined $\ols\Delta_a\equiv\ols\omega_p - \omega_a$ as an average relative atomic detuning with respect to the average pump frequency $\ols\omega_p = \sum\omega_{p_j}/N_c$, $\Delta_{c_{j}}\equiv\omega_{p_{j}} - \omega_j$ stands for the relative frequency between a pump laser and the corresponding cavity modes, and $\delta_{p_j} = \ols\omega_p - \omega_{p_j}$ is the relative detuning between an average pump frequency and pump frequency of the $j$th laser.

The corresponding many-body Hamiltonian takes the form,
\begin{equation} \label{app_mb} 
    \hat{\tilde{H}} = \int \hat\Psi^\dagger (x) \tilde{\mathcal{M}}^{(1)} \hat\Psi (x)dx + {\hat H}_{\text{int}} - \hbar\sum_j\Delta_{c_j}\big({\hat a}^\dagger_{j_+}{\hat a}_{j_+} + {\hat a}^\dagger_{j_-}{\hat a}_{j_-}\big), 
\end{equation}
where $\hat\Psi(x) = (\hat\psi_e, \hat\psi_g)^\top$ is the bosonic annihilation operator for the spinor atomic fields satisfying the usual bosonic commutation relation $[\hat\psi_{\tau}(x),\hat\psi^\dagger_{\tau'}(x')] = \delta_{\tau\tau'}\delta(x - x')$ with $\tau = \{g,e\}$, and $\tilde{\mathcal{M}}^{(1)}$ is the matrix form of the Hamiltonian density $\hat{\tilde{\mathcal{H}}}^{(1)}$, Eq.~\eqref{app_h}, except for the free Hamiltonian of the cavity fields. The second term ${\hat H}_{\text{int}}$ accounts for the two-body interaction between atoms, that for dilute quantum gases at low temperatures is well represented by local contact interactions:
\begin{equation}
    {\hat H}_{\text{int}} = \sum_\tau\frac{1}{2}g_{\tau\tau}\int\hat\psi^\dagger_{\tau}(x)\hat\psi^\dagger_{\tau}(x)\hat\psi_{\tau}(x)\hat\psi_{\tau}(x) dx + g_{eg}\int\hat\psi^\dagger_e(x)\hat\psi^\dagger_g(x)\hat\psi_g(x)\hat\psi_e(x) dx,
\end{equation}
where the two-body contact-interaction strengths $g_{gg},g_{ee}$ and $g_{eg} = g_{ge}$ are related to the respective $s$-wave scattering lengths $a^s_{\tau\tau'}$ by $g_{\tau\tau'} = 4\pi a^s_{\tau\tau'}\hbar^2/M$.

The Heisenberg equations of motion of the photonic and atomic field operators can be obtained using the many-body Hamiltonian~\eqref{app_mb}
\begin{align} \label{Heis_sys_with_e}
    i\hbar\frac{\partial {\hat a}_{j_\pm}}{\partial t} & = -i\hbar\kappa_j{\hat a}_{j_\pm} - \hbar\Delta_{c_j}{\hat a}_{j_\pm} + \hbar e^{-i\delta_{p_j} t}\int\mathcal{G}^*_j e^{\mp i k_j x}\hat\psi^\dagger_g(x)\hat\psi_e(x) dx,\nonumber\\
    i\hbar\frac{\partial\hat\psi_g (x)}{\partial t} & = \Bigg[-\frac{\hbar^2}{2M}\frac{\partial^2}{\partial x^2} + g_{gg}\hat\psi^\dagger_g (x)\hat\psi_g (x) + g_{ge}\hat\psi^\dagger_e (x)\hat\psi_e (x)\Bigg]\hat\psi_g (x)\nonumber\\
    &+ \hbar\sum_j\big[\Omega_j^* + \mathcal{G}^*_j (\hat{a}^\dagger_{j_+}e^{-i k_j x} + \hat{a}^\dagger_{j_-}e^{i k_j x})\big]e^{-i\delta_{p_j} t}\hat\psi_e (x),\nonumber\\
    i\hbar\frac{\partial\hat\psi_e (x)}{\partial t} & = \Bigg[-\frac{\hbar^2}{2M}\frac{\partial^2}{\partial x^2} - \hbar\ols\Delta_a + g_{ee}\hat\psi^\dagger_e (x)\hat\psi_e (x) + g_{eg}\hat\psi^\dagger_g (x)\hat\psi_g (x)\Bigg]\hat\psi_e (x)\nonumber\\
    &+ \hbar\sum_j\big[\Omega_j + \mathcal{G}_j(\hat{a}_{j_+}e^{i k_j x} + \hat{a}_{j_-}e^{- i k_j x})\big]e^{i\delta_{p_j} t}\hat\psi_g (x), 
\end{align}
where we have phenomenologically included the decay of the cavity mode $-i\hbar\kappa{\hat a}_{j_\pm}$. We also used the standard bosonic commutation relation for cavity operators $[{\hat a}_{m_\pm},{\hat a}^\dagger_{l_\pm}] = \delta_{ml}$. 

In the large atom-pumps detuning limit, so that $1/\ols\Delta_a $ is the fastest time scale in the system, the excited atomic field operator $\hat\psi_e$ quickly reaches the steady state and its dynamic can be adiabatically eliminated:
\begin{equation} \label{e_ss}
    \hat\psi_{e,ss} (x) \simeq \frac{1}{\ols\Delta_a}\sum_j\big[\Omega_j + \mathcal{G}_j(\hat{a}_{j_+}e^{i k_j x} + \hat{a}_{j_-}e^{-i k_j x})\big]e^{i\delta_{p_j} t}\hat\psi_g (x), 
\end{equation}
where we neglected the kinetic energy, and contact interactions in comparison with $\ols\Delta_a$. Without loss of generality in the following, we assume $\{\mathcal{G}_j,\Omega_j\}\in\mathbb{R}$. Substituting the steady-state atomic field operator of the excited state~\eqref{e_ss} in the Heisenberg equations of motion~\eqref{Heis_sys_with_e} and ignoring the terms $\hat\psi^\dagger_e (x)\hat\psi_e (x)\propto 1/\ols\Delta_a^2$ yields a set of effective equations for the photonic and ground-state atomic field operators,
\begin{align} \label{Heis_sys}
    i\hbar\frac{\partial {\hat a}_{j_\pm}}{\partial t} & = -i\hbar\kappa_j{\hat a}_{j_\pm} - \hbar\Delta_{c_j}{\hat a}_{j_\pm} + \hbar U_j{\hat N}{\hat a}_{j_\pm} + \hbar\int\big(\eta_j e^{\mp i k_j x} + U_j\hat{a}_{j_\mp}e^{\mp 2 i k_j x}\big)\hat n (x) dx, \nonumber\\ 
    i\hbar\frac{\partial\hat\psi (x)}{\partial t} & = \Bigg[-\frac{\hbar^2}{2M}\frac{\partial^2}{\partial x^2} + g_0\hat n (x) \Bigg]\hat\psi (x) + \hbar\sum_j\Big[\frac{\Omega_j^2}{\ols\Delta_a} + U_j\big(\hat{a}^\dagger_{j_+}\hat{a}_{j_+} + \hat{a}^\dagger_{j_-}\hat{a}_{j_-} + \hat{a}^\dagger_{j_+}\hat{a}_{j_-}e^{-2i k_j x} + \hat{a}^\dagger_{j_-}\hat{a}_{j_+}e^{2i k_j x} \big)\nonumber\\
    & + \eta_j\big(\hat{a}_{j_+}e^{i k_j x} + \hat{a}_{j_-}e^{-i k_j x} +\text{H.c.}\big)\Big]\hat\psi (x), 
\end{align}
where $\hat\psi(x,t)\equiv\hat\psi_g(x,t)$ and  $g_0 \equiv g_{gg}$ for the sake of simplicity of the notation, and $\hat{N} = \int \hat{n}(x) dx$ is the operator of number of particles, with $\hat{n}(x) = \hat\psi^\dagger(x)\hat\psi(x)$. Here we have introduced $U_j\equiv \mathcal{G}_j^2/\ols\Delta_a$ as the maximum depth of the cavity-generated optical potential per photon due to two-photon scattering between the $j$th pair of the degenerate cavity modes, with $\ols\Delta_a\equiv\sum_j\omega_{p_j}/N_c - \omega_a$ being the average relative atomic detuning. The coherent scattering of photons between the $j$th pump and the corresponding nearly-resonant $j$th pair of the cavity modes is described by an effective cavity-pump strength $\eta_j\equiv\mathcal{G}_j\Omega_j/\ols\Delta_a$. We neglected the fast oscillating terms with the frequency $\delta_i-\delta_{p_j} = \omega_{p_{j}} - \omega_{p_{i}}$ assuming that the frequency difference between pumps is large. The effective many-body Hamiltonian of the system which reproduces the coupled, nonlinear dynamics of the system can be read out from equations~\eqref{Heis_sys} and is given in Eq.~1 in the main text. In the following to keep the number of variables in a reasonable amount we set $\eta_j = \eta, U_j = U, \kappa_j = \kappa$.

\subsection{Cavity-mediated effective interaction potential}

In the limit of large cavity detunings $|\Delta_{c_j}|$ and/or large cavity-field decay rates $\kappa$ cavity modes evolve much faster than the center-of-mass motion of the atoms and quickly attain their steady state, so we can set $\partial{\hat a}_{j_\pm}/\partial t = 0$ in the first equation of~\eqref{Heis_sys},
\begin{equation} \label{a_ss}
{\hat a}_{j_\pm,\text{ss}} = \frac{1}{\tilde{\delta}_{c_j}}\Big[U\hat{a}_{j_\mp}\int e^{\mp 2 i k_j x}\hat n (x) dx + \eta\int e^{\mp i k_j x} \hat n (x) dx\Big],
\end{equation}
where $\tilde{\delta}_{c_j} \equiv \Delta_{c_j} - U \hat{N} + i\kappa$. By unraveling the connection between counterpropagating modes in Eq.~\eqref{a_ss}, one can receive the following expression for the steady state solution
\begin{equation}
{\hat a}_{j_\pm,\text{ss}} = \frac{\eta}{\tilde{\delta}_{c_j}}\int e^{\mp i k_j x} \hat n (x) dx + \mathcal{O}\Big(\frac{1}{\ols\Delta_a^2}\Big),
\end{equation} 
which can be substituted into the Heisenberg equation of motion for the atomic field operator~\eqref{Heis_sys}, to yield an effective atomic Hamiltonian
\begin{equation}
{\hat H}_\text{eff-at} = \int\hat\psi^\dagger(x)\Big[-\frac{\hbar^2}{2M}\frac{\partial^2}{\partial x^2} + {g_0} \hat{n}(x)\Big]\hat\psi(x)dx + \iint\mathcal{D}(x,x')\hat{n}(x)\hat{n}(x')dxdx' + \mathcal{O}\Big(\frac{1}{\ols\Delta_a^3}\Big),
\end{equation} 
where the cavity photons mediated potential is given by
\begin{equation}
\mathcal{D}(x,x') = \sum_j\frac{4\eta^2 \text{Re}(\tilde{\delta}_{c_j})}{|\tilde{\delta}_{c_j}|^2}\cos{[k_j(x-x')]}. 
\end{equation} 

The cavity-mediated interaction potential can be further shaped, to resemble for example the 1D electric dipole-dipole potential, by adjusting the parameters of the external pump lasers. As was shown in Ref.~\cite{Holzmann2021}, the variations in pump-laser parameters, such as frequencies and intensities, allowed the simulation of Coulomb interactions between particles trapped along an optical nanofiber.

\subsection{Density-contrast plot}

\begin{figure}[t!]
  \begin{minipage}[c]{0.5\textwidth}
    \includegraphics[width=0.7\textwidth]{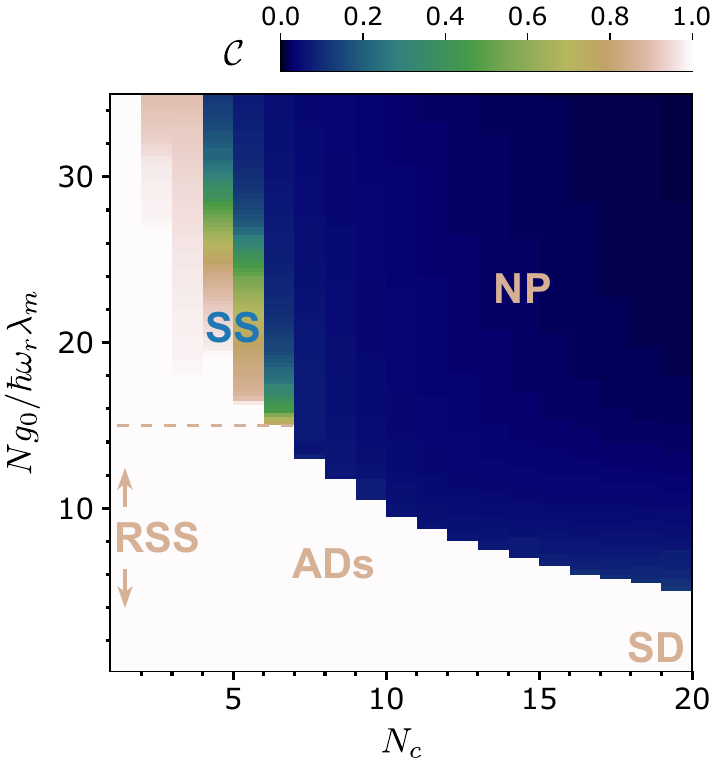}
  \end{minipage}\hfill
  \begin{minipage}[c]{0.5\textwidth}
    \caption{The density contrast ($\mathcal{C}$) as a function of the contact-interaction strength $N g_0/\hbar\omega_r\lambda_m$ and the number of involved modes $N_c$. When $\mathcal{C} = 0$ the steady state is the NP corresponding to the uniform atomic BEC and vacuum radian fields, while nonzero values of $\mathcal{C}$ signal superradiant crystalline states. The other parameters are the same as in Fig. 2 in the main text.}
    \label{fig:Density_contrast}
  \end{minipage}
\end{figure}

To compliment the phase diagram of the system in the main text, see Fig.~2, we provide here an additional plot for the density contrast $\mathcal{C}$ to characterize the existence of crystalline order 
\begin{equation}
  \mathcal{C} = \frac{n_{\text{max}} - n_{\text{min}}}{n_{\text{max}} + n_{\text{min}}}, 
\end{equation}
where $n_{\text{max}}$ ($n_{\text{min}}$) is the maximum (minimum) of the atomic density in the bulk. When $\mathcal{C} = 0$ the steady state is the NP corresponding to the uniform atomic BEC and vacuum radian fields, while nonzero values of $\mathcal{C}$ signal superradiant crystalline states.

The density contrast $\mathcal{C}$ is depicted in Fig.~\ref{fig:Density_contrast} in the same parameter regime of Fig.~2 in the main text. Note that although the density contrast, Fig.~\ref{fig:Density_contrast}, seems to be similar to the superfluid fraction, Fig~2(a), with only its color filliped, they represent very different physical quantities. The former characterizes the crystalline order and quantifies the broken translational symmetry along $x$ due to the density modulation, and the latter characterizes the superfluid fraction of a state and is associated with a broken global $U(1)$ symmetry. Although it is true that by increasing density contrast (i.e., density modulation), the superfluid fraction decreases, there is, however, no explicit relation between these two quantities.

\subsection{Nature of phase transitions}

\begin{figure}[h!]
\centering
\includegraphics [width=0.98\linewidth]{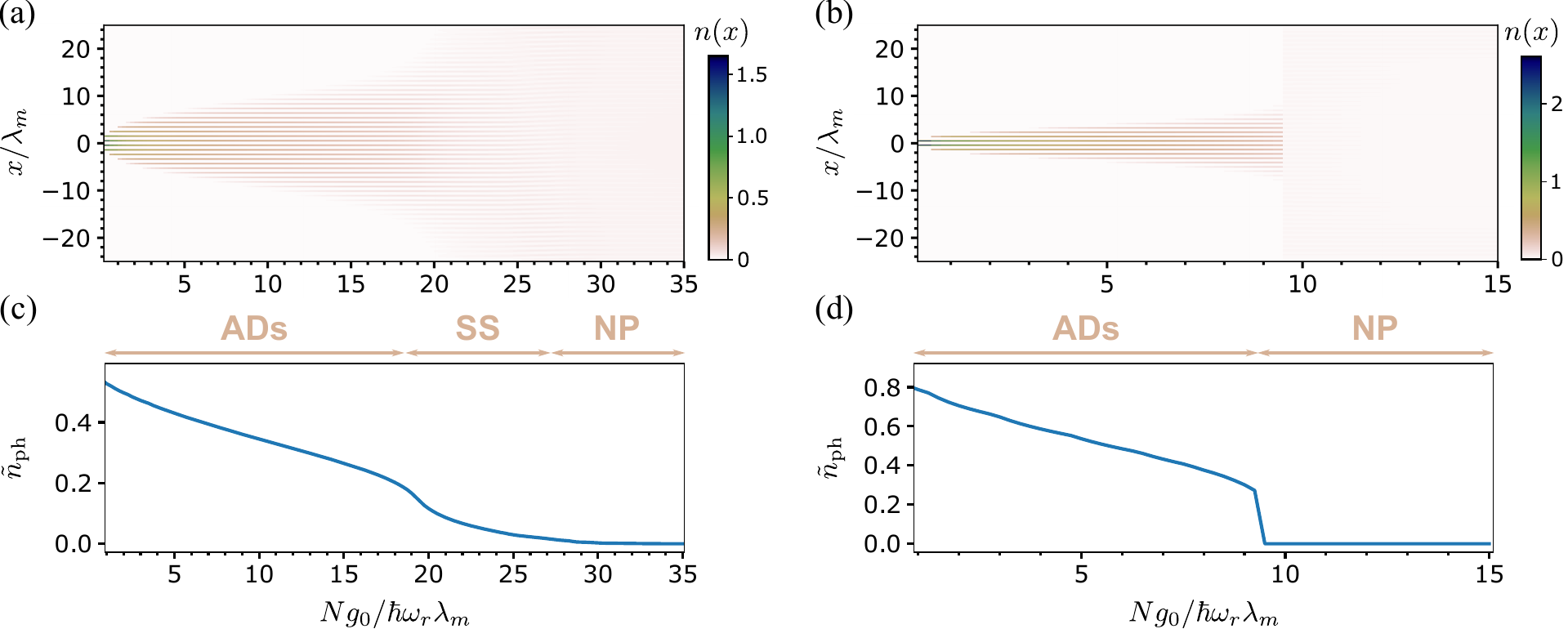}
\caption{Vertical cuts through the phase diagrams of Figs.~2(a) and (b) in the main text at $N_c = 4$ (a) and (c), and at $N_c = 10$ (b) and (d). The upper (lower) row shows the corresponding change in the atomic density $n(x)$ (the rescaled total photon number $\tilde{n}_{\rm ph}$) for a varying strength of two-body contact interactions $N g_0/\hbar\omega_r\lambda_m$. The second-order phase transition between NP and SS is smeared out due to the box potential (i.e., the hard-wall boundary condition) and consequent residual photon scattering from the BEC edges, while the  transition between NP and ADs is evidently first order.}
\label{fig:SM_cut_g}
\end{figure}

\begin{figure}[h!]
\centering
\includegraphics [width=0.98\linewidth]{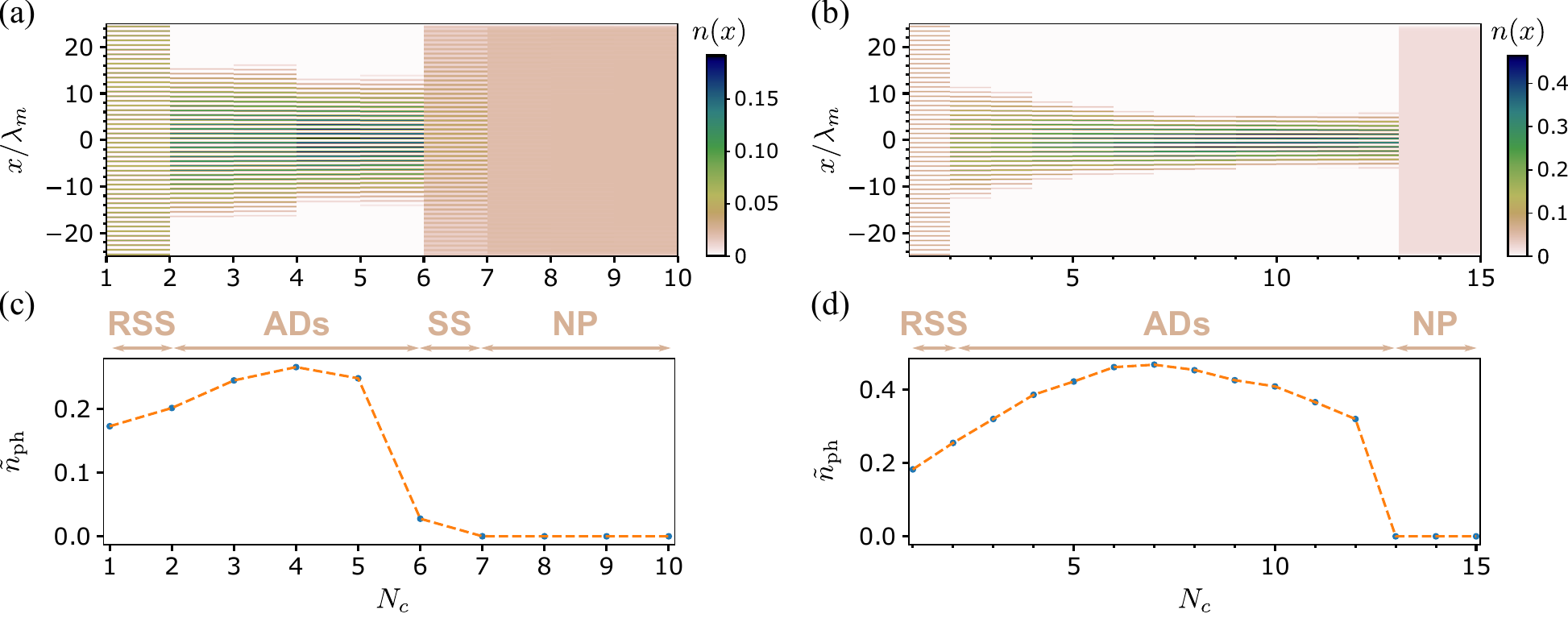}
\caption{Horizontal cuts through the phase diagrams of Figs.~2(a) and (b) in the main text at $N g_0 = 15\,\hbar\omega_r\lambda_m$ (a) and (c), and at $N g_0 = 7.5\,\hbar\omega_r\lambda_m$ (b) and (d). The upper (lower) row shows the corresponding change in the atomic density $n(x)$ (the rescaled total photon number $\tilde{n}_{\rm ph}$) for a varying number of involved modes $N_c$.}
\label{fig:SM_cut_N}
\end{figure}

To illustrate the character of phase transitions more clearly we show in Figs.~\ref{fig:SM_cut_g} and~\ref{fig:SM_cut_N} vertical and horizontal cuts, respectively, from the phase diagrams in the main text; Fig.~2(a,b). As shown in Fig.~\ref{fig:SM_cut_g}, by increasing the strength of the collisional interactions $N g_0/\hbar\omega_r\lambda_m$ the system undergoes a phase transition from ADs to NP, while for small number of involved modes $N_c$ the phase transition occurs via the intermediate SS phase, see Fig.~\ref{fig:SM_cut_g}(a) and (c). The phase transition from ADs to NP is first order, as the total photon number exhibits a discontinuous behavior [Fig.~\ref{fig:SM_cut_g}(d)], while the phase transition from SS to NP is the second order [Fig.~\ref{fig:SM_cut_g}(c)], however due to the finite size effects the sharp transition to the zero photon number is missing. On the other hand, all superradiant states cross over smoothly to one another.  

The similar situation is observed by varying the number of the involved modes $N_c$, see Fig.~\ref{fig:SM_cut_N}. In this case the SS phase appears for the stronger two-body contact interactions. 

\subsection{Additional phase diagrams for fixed number of pumped modes $N_c$}

\begin{figure}[t!]
\centering
\includegraphics [width=0.98\linewidth]{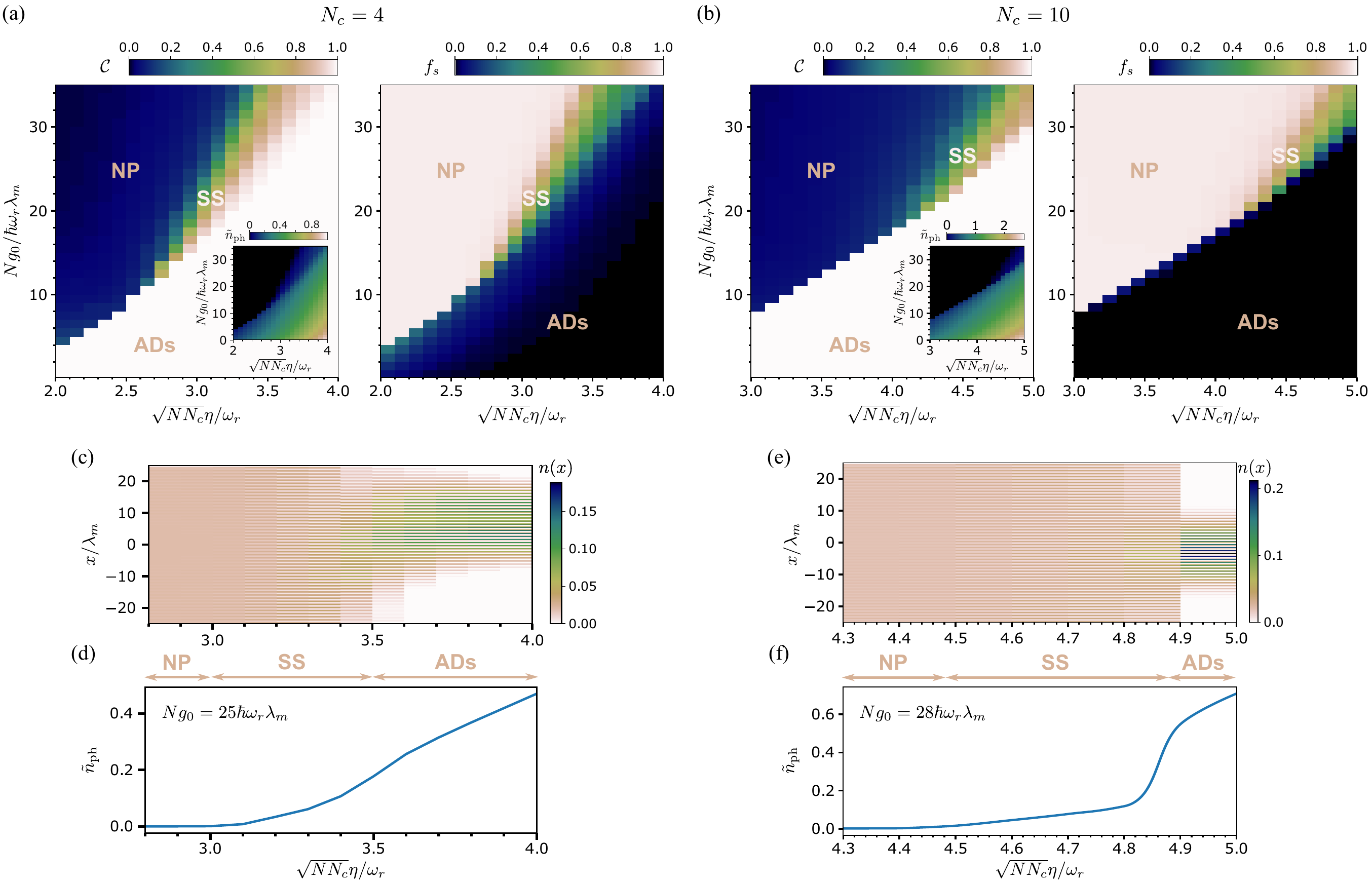}
\caption{Additional phase diagrams of the system for a fixed number of involved modes. The density contrast $\mathcal{C}$ and the superfluid fraction $f_s$ as functions of the contact-interaction strength $N g_0/\hbar\omega_r\lambda_m$ and the effective pump strength $\sqrt{N N_c}\eta/\omega_r$ for fixed number of involved modes $N_c = 4$ (a) and $N_c = 10$ (b). Note how the boundaries are shifted in the two cases $N_c = 4$ and 10. In particular, SS moves to higher interactions and pump strengths. The insets in (a) and (b) show the rescaled total photon number $\tilde{n}_{\rm ph}$ in all modes in the same parameter plane. The other parameters are the same as in Fig. 2 in the main text: $(N U,\kappa) = (-1,1)\,\omega_r$ and $\lambda_m = L_{\text{cav}}/m$ with $m = 50$ being the lowest mode in the series. (c) - (f) Horizontal cut through the phase diagrams of Figs.~\ref{fig:Add_ph_diag_Nc_const}(a,b) for $N_c = 4$ at $N g_0 = 25\,\hbar\omega_r\lambda_m$ [(c),(d)] and for $N_c = 10$ at $N g_0 = 28\,\hbar\omega_r\lambda_m$ [(e),(f)]. The upper (lower) row shows the corresponding change in the atomic density $n(x)$ (the rescaled total photon number $\tilde{n}_{\rm ph}$) for a varying strength of pumps $\sqrt{N N_c}\eta/\omega_r$. }
\label{fig:Add_ph_diag_Nc_const}
\end{figure}

In the main text, we provided the phase diagrams in the $\{N_c,\:N g_0/\hbar\omega_r\lambda_m\}$ plane, since by changing $N_c$ one modifies the \emph{range} of interactions, see Fig.~1 in the main text. However, phase diagrams in the plane of $\{\sqrt{N N_c}\eta/\omega_r,\:N g_0/\hbar\omega_r\lambda_m\}$ for fixed $N_c$ can provide another complementary perspective, as by changing the strength of the pump $\eta$ one alters the \emph{depth} of the cavity potential. The phase diagrams are presented in Fig.~\ref{fig:Add_ph_diag_Nc_const}(a,b) for $N_c = 4$ (a) and $N_c = 10$ (b). The resultant phases (NP, SS, and ADs) can be clearly distinguished from each other in these plots. The RSS and SD phases do not appear on the presented phase diagrams, since the RSS phase reveals itself mostly for $N_c = 1$, while the SD phase can still arise for given numbers of modes, but needs stronger pumps. Based on this, we do not present additional plots for IPR here, since this quantity can only help to distinguish RSS and SD phases, which are missing here.

Let us note how the phase boundaries change for different values of $N_c$ in the phase diagram of Fig.~\ref{fig:Add_ph_diag_Nc_const}(a,b). In particular, SS moves to higher interactions and pump strengths for larger $N_c$. For example, although the SS phase does not appear for $N_c = 10$ and $\sqrt{N N_c}\eta = 3.2\,\omega_r$ [corresponding to $N_c=10$ cut through the phase diagrams given in the main text, see Fig~\ref{fig:SM_cut_g}(b,d) in the SM], by increasing the strength of the pump lasers the SS emerges as shown clearly in Fig.~\ref{fig:Add_ph_diag_Nc_const}(b).

As it was noticed before, all the superradiant phase cross over smoothly to one another. This is also the case for fixed $N_c$ and changing $\eta$ as shown in Fig.~\ref{fig:Add_ph_diag_Nc_const}(c-f) here. As one can see, there is no sharp phase boundary between superradiant states even in the plane of $\{\sqrt{N N_c}\eta/\omega_r,\:N g_0/\hbar\omega_r\lambda_m\}$. The second-order phase transition between NP and SS is again smeared out due to the box potential (i.e., the hard-wall boundary condition) and consequent residual photon scattering from the BEC edges.

\subsection{Linearized equations}

To find the collective excitation spectrum we linearize the Heisenberg equations of motion given in the main text Eq.~(4) around the mean field stationary-state solutions $\psi_0 (x)$ and $\alpha_{0,j_\pm}$. We consider the deviations from the stationary solution in the form $\psi(x,t) = e^{-i\mu_0 t/\hbar}[\psi_0(x) + \delta\psi(x,t)]$, where $\mu_0$ is the corresponding chemical potential, and $\alpha_{j_\pm}(t) = \alpha_{0,j_\pm} + \delta\alpha_{j_\pm}(t)$. To obtain the linearized equations we keep only linear terms in $\delta\psi$ and $\delta\alpha_{j_\pm}$
\begin{align} \label{SM:lin_eqs}
   i\frac{\partial}{\partial t} \delta\alpha_{j_\pm} & = -\tilde{\delta}_{c_j}\alpha_{j_\pm} + U\mathcal{N}_{\pm 2}^{j,(0)}\alpha_{j_\mp} + \int A_{\pm}^j(\psi_0^*\delta\psi+\psi_0\delta\psi^*)dx,\nonumber\\
    i\frac{\partial}{\partial t} \delta\psi & = \frac{1}{\hbar}\big[\mathcal{H}^{(1)}_{\text{eff},0} + g_0 |\psi_0|^2- \mu_0\big]\delta\psi +\psi_0\sum_j \left(A_+^{j *}\delta\alpha_{j +} + A_+^j\delta\alpha_{j +}^* + A_-^{j *}\delta\alpha_{j -} + A_-^j\delta\alpha_{j -}^* \right) + g_0\psi_0^2\delta\psi^*,
\end{align}
where $\mathcal{N}_{\pm 2}^{j,(0)} = \int n_0(x)e^{\mp 2 i k_j x}dx$ and $\mathcal{H}^{(1)}_{\text{eff},0}$ corresponds to the effective single-particle Hamiltonian in the main text in Eq.~(2) with inserted mean-field solutions. We also have defined $A_{\pm}^j \equiv U(\alpha_{0,j_\pm} + \alpha_{0,j_\mp}e^{\mp 2 i k_j x}) + \eta e^{\mp i k_j x}$ for shorthands. As linearized equations~\eqref{SM:lin_eqs} couple the deviations $\delta\psi$ and $\delta\alpha_{j\pm}$ to their complex conjugates, we search the solution in the form $\delta\psi(x,t)=\delta\psi^{(+)}(x)e^{-i\omega t}+[\delta\psi^{(-)}(x)]^*e^{i\omega^* t}$ and $\delta\alpha_{j\pm}(t)=\delta\alpha_{j\pm}^{(+)}e^{-i\omega t}+[\delta\alpha_{j\pm}^{(-)}]^*e^{i\omega^* t}$. Substituting these ansatz in Eq.~\eqref{SM:lin_eqs} and writing down separately the equations for the positive- and negative-frequency components of the quantum fluctuations yields a set of coupled Bogoliubov-type equations,  
\begin{align} \label{lin_eqs_pos_neg}
\omega\delta\alpha_{j\pm}^{(+)}&= -\tilde{\delta}_{c_j}\delta\alpha_{j\pm}^{(+)}
+U\mathcal{N}_{\pm2}^{j,(0)}\delta\alpha_{j,\mp}^{(+)}
+\int A_{\pm}^j\left[\psi_0^*\delta\psi^{(+)}+\psi_0\delta\psi^{(-)}\right]dx,\nonumber\\
\omega\delta\alpha_{j\pm}^{(-)}&= \tilde{\delta}_{c_j}^*\delta\alpha_{j\pm}^{(-)} -U\mathcal{N}_{\pm2}^{j,(0)*}\delta\alpha_{j\mp}^{(-)}
-\int A_{\pm}^{j*}\left[\psi_0^*\delta\psi^{(+)}+\psi_0\delta\psi^{(-)}\right]dx,\nonumber\\
\omega\delta\psi^{(+)}&= \frac{1}{\hbar}\left[\mathcal{H}_{\rm eff,0}^{(1)} + g_0|\psi_0|^2 - \mu_0\right]\delta\psi^{(+)} +\psi_0\sum_j \left[A_+^{j*}\delta\alpha_{j+}^{(+)} + A_+^j\delta\alpha_{j+}^{(-)} + A_-^{j*}\delta\alpha_{j-}^{(+)} + A_-^j\delta\alpha_{j-}^{(-)}\right] + 
g_0\psi_0^2\delta\psi^{(-)},
\nonumber\\
\omega\delta\psi^{(-)}&= -\frac{1}{\hbar}\left[\mathcal{H}_{\rm eff, 0}^{(1)}+ g_0|\psi_0|^2-\mu_0\right]\delta\psi^{(-)} -\psi_0^* \sum_j \left[A_+^{j*}\delta\alpha_{j+}^{(+)} + A_+^j\delta\alpha_{j+}^{(-)} + A_-^{j*}\delta\alpha_{j-}^{(+)} + A_-^j\delta\alpha_{j-}^{(-)}\right] -
g_0\psi_0^{2*}\delta\psi^{(+)}.
\end{align}
This set of equations can be written in a matrix form $\omega \mathbf{f} = \mathbf{M}_B \mathbf{f}$ with $\mathbf{f} = (\delta\psi^{(+)},\delta\psi^{(-)},\delta\alpha_{m,+}^{(+)},\delta\alpha_{m,+}^{(-)},\delta\alpha_{m,-}^{(+)},\delta\alpha_{m,-}^{(-)},\\ \ldots,\delta\alpha_{m + N_c - 1,+}^{(+)},\delta\alpha_{m + N_c - 1,+}^{(-)},\delta\alpha_{m + N_c - 1,-}^{(+)},\delta\alpha_{m + N_c - 1,-}^{(-)})^\mathsf{T}$, where the Bogoliubov matrix has the following block structure 
\begin{equation}
\mathbf{M}_{\rm B}=
\left( \begin{array}{@{}c|c@{}cc}
    M_{\delta\psi\delta\psi} & M_{\delta\psi\delta\alpha}^m & \ldots & M_{\delta\psi\delta\alpha}^{m + N_c - 1} \\
   \cmidrule[0.4pt]{1-4}
    M_{\delta\alpha\delta\psi}^m & M_{\delta\alpha\delta\alpha}^m & \; & \; \\
    \vdots & \; & \ddots & \; \\
    M_{\delta\alpha\delta\psi}^{m + N_c - 1} & \; & \; & M_{\delta\alpha\delta\alpha}^{m + N_c - 1} \\
\end{array} \right).
\end{equation}
Here $M_{\delta\psi\delta\psi}$ is a $2\times 2$ submatrix, describing the connection between fluctuations of the wave function,
\begin{equation}
M_{\delta\psi\delta\psi}=
\left(   
   \begin{matrix}
      \frac{1}{\hbar}(\mathcal{H}_{\rm eff, 0}^{(1)}+ g_0|\psi_0|^2-\mu_0) & g_0\psi_0^2 \\
      -g_0\psi_0^{2*} & -\frac{1}{\hbar}(\mathcal{H}_{\rm eff, 0}^{(1)}+ g_0|\psi_0|^2-\mu_0) \\
   \end{matrix}  \\
\right). 
\end{equation}
In its turn, $M_{\delta\psi\delta\alpha}^m\;\ldots\; M_{\delta\psi\delta\alpha}^{m + N_c - 1}$ shows how fluctuations of cavity modes affects wave function fluctuations and has a form of $2\times 4 N_c$  submatrix, where
\begin{equation}
M_{\delta\psi\delta\alpha}^j=
\left(   
   \begin{matrix}
      \psi_0 A_+^{j*} & \psi_0 A^j_+ & \psi_0 A^{j*}_- & \psi_0 A^j_-  \\
      -\psi_0^* A_+^{j*} & -\psi_0^* A^j_+ & -\psi_0^* A^{j*}_- & -\psi_0^* A^j_- \\
   \end{matrix}  \\
\right).
\end{equation}
Similarly, the $4 N_c\times 2$ submatrix, consisting of corresponding blocks,
\begin{equation}
M_{\delta\alpha\delta\psi}^j=
\left(   
   \begin{matrix}
       \mathcal{I}_{+*}^j & \mathcal{I}_{+}^j \\
       -\mathcal{I}_{+}^{j*} & -\mathcal{I}_{+*}^{j*} \\
       \mathcal{I}_{-*}^j & \mathcal{I}_{-}^j \\
       -\mathcal{I}_{-}^{j*} & -\mathcal{I}_{-*}^{j*}
     \\
   \end{matrix}  \\
\right), 
\end{equation}
characterizes the influence of wave function fluctuations on cavity modes. Here we introduced the integral operators,
\begin{align}
\mathcal{I}_{\pm}^j\xi&=\int A_\pm^j(x) \psi_0(x)\xi dx,
\nonumber\\
\mathcal{I}_{\pm*}^j\xi&=\int A_\pm^j(x) \psi_0^*(x)\xi dx. 
\end{align} 
Finally, the $4 N_c\times 4 N_c$ submatrix, determining the influence of cavity fields fluctuations on itself, has a following nonzero blocks 
\begin{equation}
M_{\delta\alpha\delta\alpha}^j=
\left(   
   \begin{matrix}
       -\tilde{\delta}_{c_j} & 0 & U\mathcal{N}_{+2}^{j,(0)} & 0 \\
       0 & \tilde{\delta}_{c_j}^* & 0 & -U\mathcal{N}_{+2}^{j,(0)*}\\
       U\mathcal{N}_{-2}^{j,(0)} & 0 & -\tilde{\delta}_{c_j} & 0 \\
       0 & -U\mathcal{N}_{-2}^{j,(0)*} & 0 & \tilde{\delta}_{c_j}^*\\
   \end{matrix}  \\
\right), 
\end{equation}

To find the collective excitation spectrum we numerically find eigenvalues $\omega$ of the Bogoliubov matrix $\mathbf{M}_B$. 

\subsection{Experimental considerations}

\begin{figure}[h!]
\centering
\includegraphics [width=0.5\linewidth]{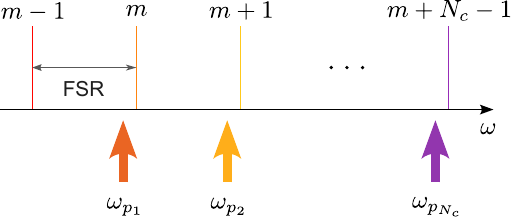}
\caption{Schematic representation of the cavity spectra for longitudinal modes. The colored lines represent longitudinal modes with $\omega_j = 2\pi c j/L_{\text{cav}}$, where $j$ is an index of the mode , $j = m - 1,\dots,m + N_c - 1$. The colored arrows denote pump lasers for every mode with index $j$ lying in the interval from $m$ to $m + N_c - 1$.}
\label{fig:modes_structure}
\end{figure}

Before moving on to a specific experimental setup, let us reveal possible restrictions that directly follow from the considered system in this paper, see Fig.~\ref{fig:modes_structure}. In order to couple many modes to the atomic transition and still stay in the dispersive regime, a free spectral range (FSR) of the cavity should be in the GHz range or smaller. This poses a constraint on the length of the cavity $L_{\text{cav}}$, which turns out to be in the order of a centimeter or larger. Inserting this condition to the formula $L = m\lambda_m$, where $m$ is the index of the lowest pumped mode in the series, and taking into account that a wavelength of the first mode is an optical wavelength, that is $\lambda_m \sim 1\mu m$ or smaller, gives a lower boundary for the index of the first mode: $m \geq 10^4$.

In the main text, we restrict ourselves to a toy situation with $m = 50$, however, this does not limit the generality of our results and findings. Such a choice of $m$ is justified by the numerical limitations, which dictate the size of the system (i.e., the number of considered unit cells $n_{\text{uc}}$) we can model. This, in turn, sets the resolution in momentum space, as $\Delta k_p = 1/n_{\text{uc}}$. On the other hand, the difference between wave vectors of neighboring cavity modes is determined by $\Delta k = 1/m$, as $\lambda_m = L_{\text{cav}}/m$. To capture atomic dynamics effectively, it is essential that the resolution in momentum space $\Delta k_p$ should be smaller or equal to $\Delta k$, which gives the condition $n_{\text{uc}} \geq m$. To this end, we chose $m = 50$, though not the most experimentally relevant value, which does encapsulate the significant phenomena we aim to highlight.

\begin{figure}[h!]
\centering
\includegraphics [width=0.5\linewidth]{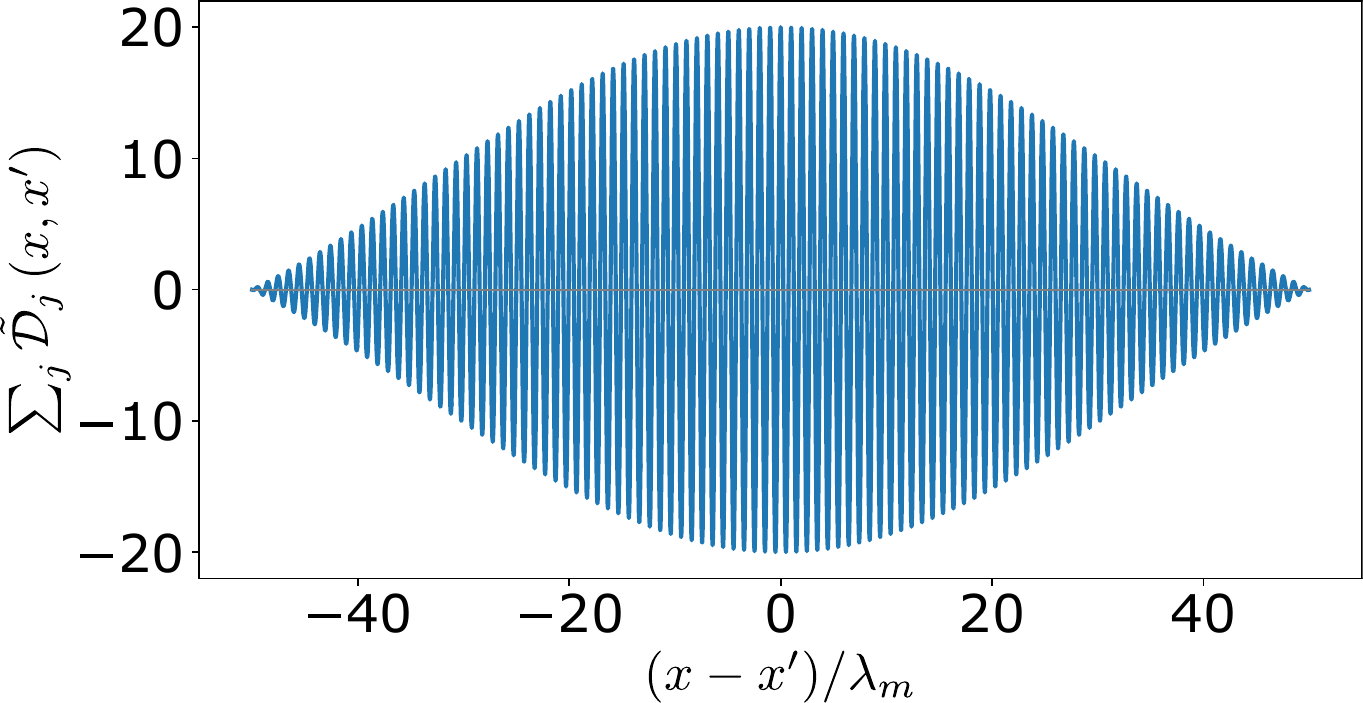}
\caption{The dimensionless cavity-mediated interaction potential $\sum_j\mathcal{\tilde{D}}_j(x,x')$ as a function of the relative interatomic distance for $N_c = 20$ pumped modes with a frequency comb with a repetition rate $\approx 0.3$ THz and FSR of the cavity is around 10 MHz. The dimensionless interatomic distance $x - x'$ is given in the units of $\lambda_m\approx 1\mu$m and therefore the effect of the short-range cavity-mediated interactions is visible on the size of a BEC.}
\label{fig:SM_optical_potential}
\end{figure}

Let us now give a more concrete example. In the recent experiment~\cite{Johnson2019} an optical fiber ring resonator with adjustable length was used for the observation of strong coupling between the cavity mode and an ensemble of Cs atoms. The length of the cavity was set to 30 m, which leads to $\nu_{\text{FSR}} = 7.1$ MHz, and the wavelength of the cavity mode was locked by a probe laser with $\lambda = 852$ nm. This corresponds to a mode number $m\approx 3\times 10^7$. Starting from this mode as in our scheme, in order to see the effect of short-range cavity-mediated interactions on the size of a BEC, one should consider a large number of pumped modes. One can significantly simplify an experimental setup by using a frequency comb~\cite{Thorpe2008}, with a repetition rate much higher than $\nu_{\text{FSR}}$, as a pump for the cavity modes~\cite{Torggler2020}. In Fig.~\ref{fig:SM_optical_potential} we show the resulting dimensionless cavity-mediated interaction potential $\sum_j\mathcal{\tilde{D}}_j(x,x')$ as a function of the relative interatomic distance for $m= 3\times 10^7$ and $N_c = 20$ pumped modes spaced by the repetition rate of the comb pumps (approximately 0.3 THz). The dimensionless interatomic distance is given in the units of $\lambda_m\approx 1\mu$m and therefore the effect of the short-range cavity-mediated interactions reveals itself on the size of a BEC. To couple the atomic transition to twenty (non-adjacent) cavity modes demands around $0.3\times 20 = 6$ THz, which can be comfortably fitted in an optical transition such as the ${}^{87}\text{Rb}$ $D_2$ line with a frequency of around $2\pi\times 384$ THz, while the frequency of the first mode is $\omega_m=2\pi\times 352$ THz.

Moreover a BEC with an extended length~\cite{Lim2021} or composed of atoms with broader optical transitions, such as ${}^{23}\text{Na}$~\cite{Davis1995}, ${}^{40}\text{Ca}$~\cite{Kraft2009} or ${}^{86}\text{Sr}$~\cite{Stellmer2009}, could provide additional flexibility for experimental realization.